\documentclass[conference]{IEEEtran}
\usepackage{graphicx}
\usepackage{amsmath}
\usepackage{lipsum}
\usepackage{amsthm}
\usepackage[strings]{underscore}
\usepackage{amsfonts}
\usepackage{caption}
\usepackage{siunitx}
\usepackage{amssymb}
\usepackage{subcaption}
\usepackage{placeins}
\usepackage{lettrine}
\usepackage{multicol, blindtext}
\usepackage{bbm}
\usepackage{algorithm}
\usepackage{algpseudocode}
\usepackage{authblk}
\usepackage{cite}

\makeatletter

\newcommand{\Rmnum}[1]{\expandafter\@slowromancap\romannumeral #1@}
\newenvironment{psmallmatrix}
  {\left[\begin{smallmatrix}}
  {\end{smallmatrix}\right]}
\newtheorem{theorem}{Theorem}

\makeatletter
\def\BState{\State\hskip-\ALG@thistlm}
\makeatother
\ifCLASSINFOpdf
\else
\fi

\begin{document}
%
\title{Minimizing The Age of Information: NOMA or OMA?}
\author[*]{Ali Maatouk}
\author[*]{Mohamad Assaad}
\author[$\dagger$]{Anthony Ephremides}
\affil[*]{TCL Chair on 5G, Laboratoire des Signaux et Syst\`emes, CentraleSup\'elec, Gif-sur-Yvette, France }
\affil[$\dagger$]{ECE Dept., University of Maryland, College Park, MD 20742}
\maketitle

\begin{abstract}
In this paper, we examine the potentials of Non-Orthogonal Multiple Access (NOMA), currently rivaling Orthogonal Multiple Access (OMA) in 3rd Generation Partnership Project (3GPP) standardization for future 5G networks Machine Type Communications (MTC), in the framework of minimizing the average Age of Information (AoI). By leveraging the notion of Stochastic Hybrid
Systems (SHS), we find the total average AoI of the network in
simple NOMA and conventional OMA environments. Armed with this, we provide a comparison between the two schemes in terms of average AoI. Interestingly, it will be shown that even when NOMA achieves better spectral efficiency in comparison to OMA, this does not necessarily translates into a lower average AoI in the network.
\end{abstract}


%
\IEEEpeerreviewmaketitle

\section{Introduction}
\lettrine{I}{}n recent years, there has been a dramatic proliferation of
research on the notion of Age of Information (AoI). The AoI was first introduced in \cite{6195689} and was motivated by a variety of new applications requiring fresh information at a monitor to accomplish specific tasks. Some non-exclusive examples include environmental monitoring and vehicular networks \cite{5597912}\cite{5307471}. In these applications, an information source generates time-stamped status updates that are sent to the monitor. The eventual goal is to make the monitor have the freshest possible knowledge about the information of interest. Although the idea appears simple, it was shown that the study of the AoI, even in the simplest scenarios, to be far from trivial.

The seminal work in \cite{6195689} has opened the gate to an extensive number of research papers on the subject of AoI. In the aforementioned reference, the new metric of AoI was investigated in the standard First-Come-First-Served (\textbf{FCFS}) discipline scenarios: M/M/1, M/D/1 and D/M/1. The ability to manage the packets of the queue was highlighted as a mean of reducing the AoI in \cite{6875100}. Up to that point, the literature was mainly focused on the case where packets arrivals are considered to be stochastic. The AoI was then investigated in the case where sources have control over the packets arrival (i.e. sampling problems). More specifically, a $``$generate-at-will$"$ model was considered in \cite{8000687} and an interesting conclusion was drawn that a zero-wait sampling policy is not always optimal in minimizing the average age. With the AoI being of interest in sensor type applications where energy is critical, the AoI has attracted attention in the case of energy harvesting sources (e.g. \cite{2018arXiv180202129A}). With much of the early work on the AoI being focused on single hop scenarios, multi-hop scenarios have been recently considered in \cite{8262777,8445981,8006593} where in the latter, it was shown that the Last-Come-First-Served (\textbf{LCFS}) discipline at relaying nodes minimizes the average age of the considered stream. As streams have normally different priorities based on the sensitivity of the information they carry, a surge in papers considering prioritized queuing can be witnessed \cite{2018arXiv180805738Z,2018arXiv180104067N,2018arXiv180511720M}. For example, the age of information in a multicast scenario where two different priority groups exist was studied in \cite{2018arXiv180805738Z}. The minimization of AoI has also been theoretically investigated in a Carrier Sense Multiple Access (\textbf{CSMA}) environment \cite{2019arXiv190100481M}.

Scheduling with the aim of minimizing the average age of a broadcast network (e.g. celullar network) has been examined extensively in the litearture (e.g. \cite{8006590}). However, the majority of the papers published in that regards focus mainly on the case where only \emph{one} user can be active on a particular resource (e.g. a subcarrier for a certain amount of time). This is in agreement with Orthogonal Multiple Access (\textbf{OMA}) schemes that have always been the norm for multiple access techniques. For instance, Time-Division Multiple Access (\textbf{TDMA}) and Code-Division Multiple
Access (\textbf{CDMA}) were respectively used in the Global System
for Mobile Communications (\textbf{GSM}) and IS-95 standards
of 2G networks. 3G networks also employed an OMA scheme, the so called Wideband CDMA (\textbf{WCDMA}). The Long-Term Evolution (\textbf{LTE}) standard uses Orthogonal Frequency-Division Multiple Access (\textbf{OFDMA}) on the downlink and
Single-Carrier Frequency-Division Multiple Access (\textbf{SC-FDMA})
on the uplink. Moreover, orthogonal multicarrier transmission technology such as Orthogonal Frequency-Division Multiplexing (\textbf{OFDM}),
is also used in WiMAX, Wi-Fi, and terrestrial digital video
broadcasting (\textbf{DVB-T}). Based on the preceding, we can see that the orthogonality of different users' signals was always perceived as
a most desirable property. However, a rival to OMA known as Non-Orthogonal Multiple Access (\textbf{NOMA}) \cite{6692652,8303689,8377142} where \emph{more than one} user transmits on a specific resource, has
been receiving increasing attention from the cellular networks research community and is now perceived as a most promising technology for
future cellular systems including 5G. More specifically, the 3rd Generation Partnership Project (\textbf{3GPP}) is considering NOMA for Machine Type Communications (\textbf{MTC}) in the next 5G networks \cite{3gpp.38.812}. The main motivation of employing NOMA comes from well-established results
in multi-user information theory which state that superposition coding along with Successive Interference Cancellation (\textbf{SIC}) provides significant improvement in spectral efficiency in comparison to OMA \cite{tse_viswanath_2005}. With
the AoI being of broad interest in machine-type applications \cite{6195689}\cite{5597912}, the potentials of NOMA in minimizing the average AoI of the network has to be investigated. This motivated our work where we take simple NOMA settings and leverage the notion of Stochastic Hybrid Systems (\textbf{SHS}) to find a closed form of the total average age of the network in this case. By using the same tools, we find a closed form of the OMA environment counterpart. Armed with these closed forms, we investigate the potentials of NOMA in terms of minimizing the average AoI of the network in comparison to OMA. We will show that even when NOMA provides spectral efficiency enhancement with respect to OMA, this does not directly translate into a better AoI performance. This is a consequence of a key difference between data communication systems and status update systems: in the first, all packets are equally important and therefore an increase in the number of delivered packets (i.e. better spectral efficiency) is always desirable. On the other hand, this is not the case in status updates where a packet's importance depends on its potentials to give fresh knowledge to the monitor. To the knowledge of
the authors, this is the first work that investigates NOMA settings in the framework of minimizing the average AoI of the network.

The paper is organized as follows: Section \Rmnum{2} describes the system model of NOMA. Section \Rmnum{3} presents the theoretical results on the total average age of the network in both NOMA and conventional OMA environments. Section \Rmnum{4} provides the numerical results that utilize the theoretical findings while Section \Rmnum{5} concludes the paper.

\section{NOMA System Model}
\begin{figure}[!ht]
\centering
\includegraphics[width=.8\linewidth]{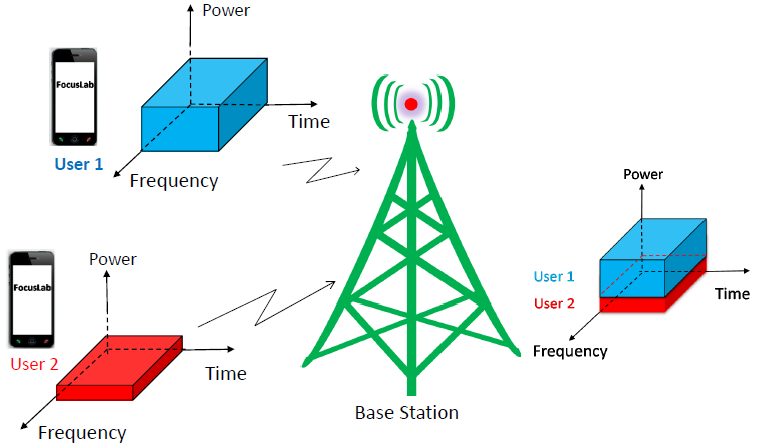}
\caption{Illustration of uplink NOMA}
\setlength{\belowcaptionskip}{-5pt}
\label{nomaup}
\end{figure}

\begin{figure}[!ht]
\centering
\includegraphics[width=.7\linewidth]{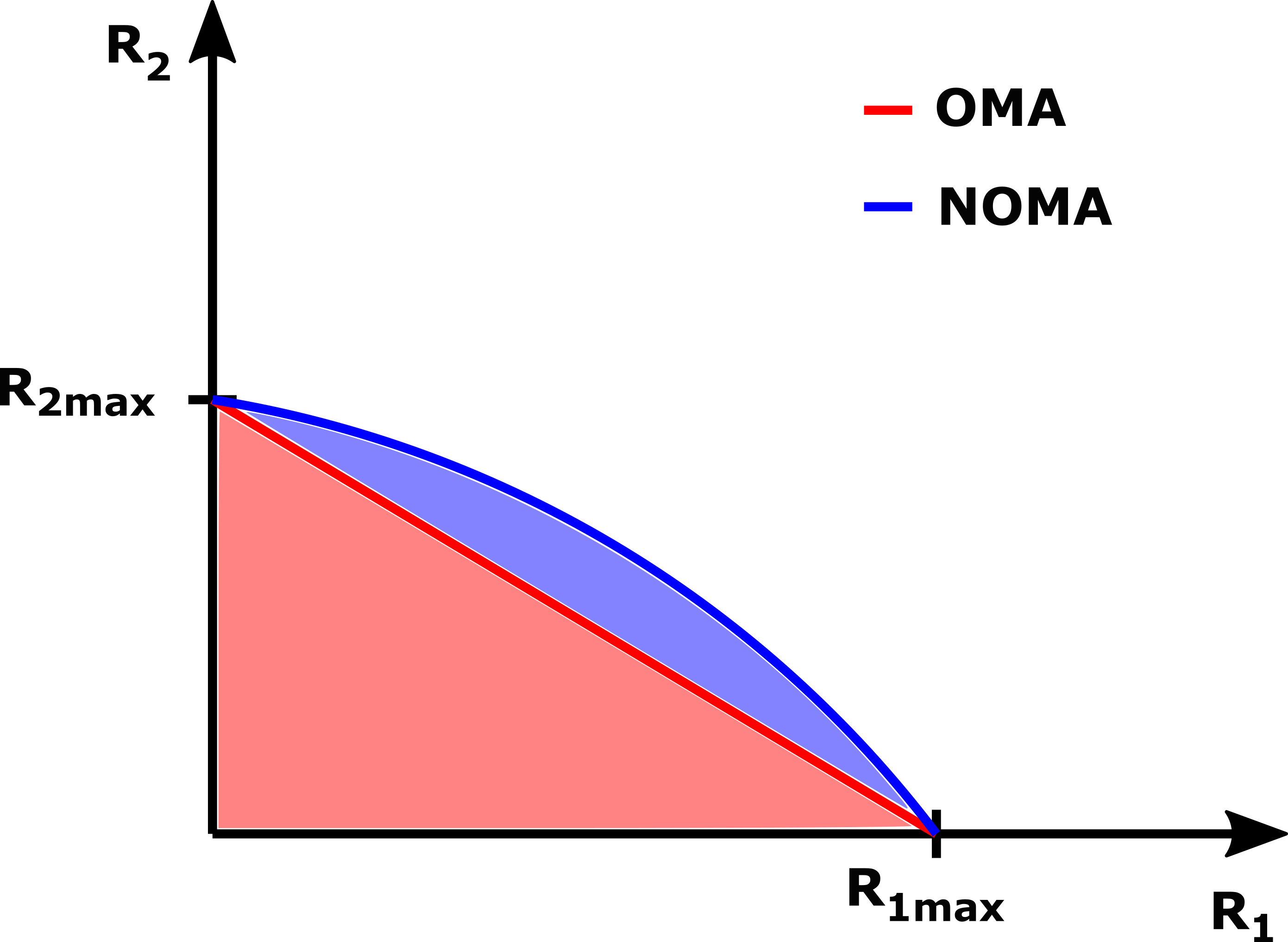}
\caption{Illustration of the achievable region in OMA vs NOMA}
\setlength{\belowcaptionskip}{-5pt}
\label{noma}
\end{figure}

In an OMA environment, each user is assigned a single orthogonal block that is free of any interference from other users (e.g. a single subcarrier for a certain duration of time). On the other hand, in NOMA, two or more users are granted access to the same orthogonal block by assigning to each of them different power levels \cite{PNOMA1}. The idea is that the different power levels will help distinguish between the users' signals and therefore an interference cancellation procedure using SIC can take place. To describe this principle, we explain the scenario reported in Fig. \ref{nomaup}. It is worth mentioning that superposing the signals of more than $2$ users on a single orthgonal block can lead to severe degradation in the Bit Error Rate (\textbf{BER}) \cite{8377142} and we therefore focus in the rest of the paper on the $2$ users case. As seen in Fig. \ref{nomaup}, User $1$ has a strong signal power $P_1$ and User $2$ has a weaker signal power $P_2$. Then, User $1$'s signal can be easily detected by the receiver simply by sending the received signal to a threshold detector directly. However, this type of detection does not apply to User $2$, because its signal is buried in the stronger User $1$'s signal. Therefore, after detecting User $1$'s signal, the receiver subtracts it from the received signal and feeds it to a threshold detector to detect the signal of User $2$. Notice that similar discussions also hold for the downlink.
\\
By employing the described scheme, the additional blue region reported in Fig. \ref{noma} can be achieved. More specifically, let us consider $R_{1max}$ and $R_{2max}$ as the maximal rates that can be achieved by users $1$ and $2$ respectively if the channel is always granted to them with no interference. Using NOMA, a higher \emph{sum-rate} (i.e. spectral efficiency) can be achieved but, due to interference, the individual rates achieved by each user are strictly smaller than $R_{1max}$ and $R_{2max}$. We refer the readers to \cite{2015arXiv150407751X} for a more elaborate discussion on the achieved region by NOMA.
\\
To properly model this scenario, we consider that the packets' arrival for each user $k \:\:\forall k\in\{1,2\}$ is exponentially distributed with a rate of $\lambda_k$. The transmission time of the packets is assumed to be exponentially distributed with an average rate of $\mu_k$ when there is no interfering user scheduled on the same orthogonal block. When the two users are superposed on the same orthogonal block, the transmission time rate of each user $k$ is $\mu'_k$ in the blue region such that:
\begin{equation}
\mu'_1(\delta)<\mu_1 \quad\quad \mu'_2(\delta)<\mu_2
\end{equation}
\begin{equation}
\mu'_1(\delta)+\mu'_2(\delta)>\delta \mu_1+ (1-\delta)\mu_2 \quad\forall\delta\in[0,1]
\end{equation}
where $\delta$ is the sharing factor of power and frequency/time in NOMA and OMA respectively \cite{2015arXiv150407751X}. We forgo queuing in this network; each user keeps at most one packet in its system. Upon a new arrival of a packet to user $k$, the packet of user $k$ that is currently available (or being served) is preempted and discarded. This is motivated by the fact that a preemptive M/M/1/1 scenario was shown to minimize the average AoI in the case of exponential transmission time \cite{8006593}. Moreover, this model is useful in realistic scenarios (e.g. congested wireless network) as the service time would be dominated by the MAC access delay \cite{2018arXiv180307993Y}. 

In this scenario, the instantaneous age of information at the receiver (monitor) of user $k$ at time instant $t$ is defined as:
\begin{equation}
\Delta_k(t)=t-U_k(t) \quad\quad k=1,2
\end{equation}
where $U_k(t)$ is the time stamp of the last successfully received packet by the receiver side of user $k$. Clearly the evolution of the age will depend on the arrival process of each user, along with the time spent to transmit the packet. The ultimate goal therefore consists of minimizing the total average age of the network defined as:
\begin{equation}
\overline{\Delta}=\sum_{k=1}^{2}\overline{\Delta}_k=\sum_{k=1}^{2}\lim_{\tau\to\infty}\frac{1}{\tau}\int_{0}^{\tau}\Delta_k(t)dt
\end{equation}
\section{Theoretical Analysis}
\subsection{Introduction to SHS}
The typical approach to calculate the average age through graphical methods, such as the one adopted in \cite{6195689}, is challenging in queuing systems where packets can be dropped (i.e lossy systems) \cite{2018arXiv180307993Y}. In our paper, we take the SHS approach to find the total average AoI of the network in question. This approach revolves around considering the states $(q(t),\boldsymbol{x}(t))$ where $q(t)\in\mathbb{Q}$ is a discrete process that captures the evolution of the network
while $\boldsymbol{x}(t)\in\mathbb{R}^{4}$ is a continuous process that represents the age process of user $1$ and $2$ at the monitor along with the packets in their system. To be more precise:
\begin{equation}
\boldsymbol{x}(t)=[x_0(t),x_1(t),x_{2}(t),x_{3}(t)]
\end{equation}
where:
\begin{equation*}
\begin{cases}
x_{2k}(t)& \text{Age of user}\: $k+1$\: \text{at the monitor} \: 0\leq k\leq 1\\
x_{2k+1}(t)& \text{Age of the packet of user} \: $k+1$ \: 0\leq k\leq 1\\
\end{cases}
\label{Stationary}
\end{equation*}
We outline the general idea of SHS in the following (we refer the readers to \cite{DBLP:journals/corr/YatesK16} for more details).

The Markov process $q(t)$ can be modeled by a Markov chain $(\mathbb{Q},\mathbb{L})$ in which each state $q\in\mathbb{Q}$ is a vertex and each transition\footnote{It is worth mentioning that unlike a typical continuous-time Markov chain, the chain $(\mathbb{Q},\mathbb{L})$ may include self-transitions where a reset of the continuous process $\boldsymbol{x}$ takes place but the discrete state remains the same.
} $l\in\mathbb{L}$ is a directed edge $(q_l,q'_{l})$ with a transition rate $\lambda^{(l)}\delta_{q_l,q(t)}$. We multiply by the Kronecker delta function to make sure that this transition $l$ can only occur when the process $q(t)=q_l$. We further define for each state $q$ the incoming and outgoing transitions sets respectively as:
\begin{equation}
\mathbb{L}'_q=\{l\in\mathbb{L}: q'_{l}=q\} \quad \mathbb{L}_q=\{l\in\mathbb{L}: q_{l}=q\}
\end{equation}
SHS revolves around the fact that any transition experienced by the discrete process $q(t)$ will cause a reset in the continuous process. To further elaborate, suppose transition $l$ takes place, the discrete process changes value to $q'_{l}$ and a fall in the continuous process $\boldsymbol{x'}=\boldsymbol{x}\boldsymbol{A_l}$ is witnessed. More precisely, $\boldsymbol{A_l}\in\mathbb{R}^{4}\times\mathbb{R}^{4}$ is called the reset map of transition $l$. As a final step, in each state $q\in\mathbb{Q}$, the continuous process evolves through the following differential equation $\dot{\boldsymbol{x}}=\boldsymbol{b}_q$ where $b^k_q$ is a binary element that is equal to $1$ if the age process $x_k$ increases at a unit rate when the system is in state $q$ and is equal to $0$ if it keeps the same value.

To calculate the average age of the system through SHS, the following quantities for each state $q\in\mathbb{Q}$ need to be defined:
\begin{equation}
\pi_{q}(t)=\mathbb{E}[\delta_{q,q(t)}]=P(q(t)=q)
\end{equation}
\begin{equation}
\boldsymbol{v}_{q}(t)=[v_{q0}(t),\ldots,v_{q4}(t)]=\mathbb{E}[\boldsymbol{x}(t)\delta_{q,q(t)}]
\end{equation} 
where $\pi_{q}(t)$ is the Markov chain's state probabilities at time instant $t$ and $\boldsymbol{v}_{q}(t)$ refers to the correlation between the age vector $\boldsymbol{x}(t)$ and the state $q$. One important consideration is that the Markov chain $q(t)$ is supposed to be ergodic and we can hence establish the existence of the steady state probability vector $\overline{\boldsymbol{\pi}}$ as the solution to the following equations:
\begin{equation}
\overline{\pi}_{q}(\sum_{l\in\mathbb{L}_q}\lambda^{(l)})=\sum_{l\in\mathbb{L}'_q}\lambda^{(l)}\overline{\pi}_{q_l} \quad q\in\mathbb{Q}
\end{equation}
\begin{equation}
\sum_{q\in\mathbb{Q}}\overline{\pi}_{q}=1
\end{equation}
As proven in \cite{DBLP:journals/corr/YatesK16}, in this case, the correlation vector $\boldsymbol{v}_{q}(t)$ converges to a limit $\overline{\boldsymbol{v}}_{q}$ such that:
\begin{equation}
\overline{\boldsymbol{v}}_{q}(\sum_{l\in\mathbb{L}_q}\lambda^{(l)})=\boldsymbol{b}_q\overline{\pi}_{q}+\sum_{l\in\mathbb{L}'_q}\lambda^{(l)}\overline{\boldsymbol{v}}_{q_l}\boldsymbol{A_l} \quad q\in\mathbb{Q}
\label{solutionv}
\end{equation}
We can therefore conclude, based on the preceding, that $\mathbb{E}[x_{2k}]=\lim\limits_{t \to +\infty} \mathbb{E}[x_{2k}(t)]=\lim\limits_{t \to +\infty}\sum\limits_{q\in\mathbb{Q}}\mathbb{E}[x_{2k}(t)\delta_{q,q(t)}]=\sum\limits_{q\in\mathbb{Q}}\overline{v}_{q2k}\:\:\forall k\in\{0,1\}$.\\
With the aforementioned results from \cite{DBLP:journals/corr/YatesK16} taken into consideration and as our goal is to find the total average AoI at the monitor of the network, we present the following theorem:
\begin{theorem}
In the case where $q(t)$ is ergodic and admits $\boldsymbol{\overline{\pi}}$ as stationary distribution, if we can find a solution for eq. (\ref{solutionv}), then the total average AoI of the network is:
\begin{equation}
\overline{\Delta}=\sum_{k=1}^{2}\overline{\Delta}_k=\sum_{q\in\mathbb{Q}}(\overline{v}_{q0}+\overline{v}_{q2})
\end{equation}
\label{theomhem}
\end{theorem}
\subsection{NOMA Scenario}
In order to simplify the average age calculations, we forgo studying the age process of both users simultaneously. Instead, we examine the network from the perspective of user $1$ solely as the network is symmetrical with respect to user $2$ by simply substituting the number $1$ with $2$ and vice-versa. We provide in the following theorem a closed form of the average age of information in the simple NOMA settings reported in Section II. The proof is based on the SHS tools explained in the previous subsection.
\begin{theorem}
In the NOMA settings, the average age of user $1$ is equal to $\overline{\Delta}_1=\sum\limits_{q\in\mathbb{Q}}\overline{v}_{q0}=\overline{v}'_1+\overline{v}'_2+\overline{v}'_4+\overline{v}'_6$ 
with $\boldsymbol{\overline{v}}'=[\overline{v}_{00},\overline{v}_{10},\overline{v}_{11},\overline{v}_{20},\overline{v}_{21},\overline{v}_{30}]^{T}$ being the solution to the linear system:
\begin{equation}
\boldsymbol{A}_2\boldsymbol{\overline{v}'}=\boldsymbol{c}_2
\label{eqlin2}
\end{equation}
where:
\begin{equation}
\boldsymbol{A}_2=
\begin{psmallmatrix}
    \lambda_1+\lambda_1 & 0 &-\mu_1 & 0 & 0  & -\mu_2 \\
     -\lambda_1 & \lambda_2+\mu_1 &0 & -\mu'_2 & 0  & 0 \\
     0 & 0 & \lambda_1+\lambda_2+\mu_1 & 0 & -\mu'_2  & 0 \\
    0 & -\lambda_2 & 0 & \mu'_1+\mu'_2 & 0  & -\lambda_1 \\
      0 & 0 &-\lambda_2 & 0 & \mu'_1+\mu'_2  & 0 \\
       -\lambda_2 & 0 &0 & 0 & -\mu'_1  & \lambda_1+\mu_2 \\
\end{psmallmatrix}
\end{equation}
\begin{equation}
\boldsymbol{c}_2=\begin{psmallmatrix}
    \overline{\pi}_0  \\
    \overline{\pi}_1  \\
\overline{\pi}_1  \\
\overline{\pi}_2 \\
\overline{\pi}_2  \\
\overline{\pi}_3 
\end{psmallmatrix}
\end{equation}
and $\overline{\boldsymbol{\pi}}$ is the stationary distribution of the Markov chain $q(t)$ that can be found by solving:
\begin{equation}
\boldsymbol{A}_1\boldsymbol{\overline{\pi}}=\boldsymbol{c}_1
\label{eqlin}
\end{equation}
where:
\begin{equation}
\boldsymbol{A}_1=
\begin{psmallmatrix}
    \lambda_1+\lambda_2 & -\mu_1 & 0 & -\mu_2 \\
    -\lambda_2 & 0 & -\mu'_1 & \lambda_1+\mu_2 \\
    0& -\lambda_2 & \mu'_1+\mu'_2 & -\lambda_1 \\
    1 & 1 & 1 & 1 
\end{psmallmatrix}
\end{equation}
\begin{equation}
\boldsymbol{c}_1=\begin{psmallmatrix}
    0  \\
    0  \\
0  \\
1  \\
\end{psmallmatrix}
\end{equation}
\label{averageagenoma}
\end{theorem}
\begin{IEEEproof}
The first step of our proof consists of defining the discrete states set $\mathbb{Q}$. In our scenario, we consider that $\mathbb{Q}=\{0,1,2,3\}$ where:
\begin{enumerate}
\item $q(t)=0$: no packets are available for either of the two users and therefore the server is idle at time $t$
\item $q(t)=1$: user $1$ is being served while user $2$ has no packet at time $t$
\item $q(t)=2$: both users are being served simultaneously at time $t$
\item $q(t)=3$: user $2$ is being served while user $1$ has no packet at time $t$
\end{enumerate}
The continuous-time state process is defined as $\boldsymbol{x}(t)=[x_{0}(t),x_{1}(t)]$ where $x_{0}(t)$ is the age of user $1$ at the monitor at time $t$ and $x_{1}(t)$ is the age of the packet in the system of user $1$ at time $t$. Our goal is therefore to use Theorem \ref{theomhem} to find  $\overline{\boldsymbol{v}}_{q}=[\overline{v}_{q0},\overline{v}_{q1}]\:\: \forall q\in\mathbb{Q}$  which will allow us to find a closed form of the average age of user $1$. We start first by summarizing the transitions between the discrete states and the reset maps they induce on the age process $\boldsymbol{x}(t)$ in the table that follows:
\begin{center}
\begin{tabular}{cccccc}
$l$ & $q_l\rightarrow q'_{l}$ & $\lambda^{(l)}$ & $\boldsymbol{xA_l}$ & $\boldsymbol{A_l}$  &  $\boldsymbol{v}_{q_l}\boldsymbol{A_l}$ \\
\hline
 $1$ & $0\rightarrow1$  & $\lambda_1$ & $[x_0,0]$ & $\begin{psmallmatrix}
    1 & 0  \\
    0& 0
\end{psmallmatrix}$  &  $[v_{00},0]$\\
 $2$ & $0\rightarrow3$  & $\lambda_2$ & $[x_0,0]$ & $\begin{psmallmatrix}
    1 & 0  \\
    0& 0
\end{psmallmatrix}$  &  $[v_{00},0]$\\
 $3$ & $1\rightarrow1$  & $\lambda_1$ & $[x_0,0]$ & $\begin{psmallmatrix}
    1 & 0  \\
    0& 0
\end{psmallmatrix}$  &  $[v_{10},0]$\\
 $4$ & $1\rightarrow0$  & $\mu_1$ & $[x_1,0]$ & $\begin{psmallmatrix}
    0 & 0  \\
    1& 0
\end{psmallmatrix}$  &  $[v_{11},0]$\\
 $5$ & $1\rightarrow2$  & $\lambda_2$ & $[x_0,x_1]$ & $\begin{psmallmatrix}
    1 & 0  \\
    0& 1
\end{psmallmatrix}$  &  $[v_{10},v_{11}]$\\
 $6$ & $2\rightarrow1$  & $\mu'_2$ & $[x_0,x_1]$ & $\begin{psmallmatrix}
    1 & 0  \\
    0& 1
\end{psmallmatrix}$  &  $[v_{20},v_{21}]$\\
$7$ & $2\rightarrow2$  & $\lambda_1$ & $[x_0,0]$ & $\begin{psmallmatrix}
    1 & 0  \\
    0& 0
\end{psmallmatrix}$  &  $[v_{20},0]$\\
 $8$ & $2\rightarrow3$  & $\mu'_1$ & $[x_1,0]$ & $\begin{psmallmatrix}
    0 & 0  \\
    1& 0
\end{psmallmatrix}$  &  $[v_{21},0]$\\
 $9$ & $3\rightarrow2$  & $\lambda_1$ & $[x_0,0]$ & $\begin{psmallmatrix}
    1 & 0  \\
    0& 0
\end{psmallmatrix}$  &  $[v_{30},0]$\\
 $10$ & $3\rightarrow0$  & $\mu_2$ & $[x_0,0]$ & $\begin{psmallmatrix}
    1 & 0  \\
    0& 0
\end{psmallmatrix}$  &  $[v_{30},0]$\\
\end{tabular}
 \captionof{table}{Stochastic Hybrid System Description}
 \label{trans}
\end{center}
In the following, we elaborate on the transitions reported in Table \ref{trans}:
\begin{enumerate}
\item The set of transitions $l=1$ and $l=9$ corresponds to the case where a new packet arrives to the empty system of user $1$. On the other hand, the set of transitions $l=3$ and $l=7$ corresponds to the case where a new packet arrival preempts the packet already in service for user $1$. In both cases, the transition will not influence the age process $x_0$ at the monitor but will reset the age of the packet of user $1$ to $0$.
\item The transition $l=2$ signals an arrival of a packet to user $2$. This transition will have no effect on the age of user $1$ at the monitor\footnote{It is worth mentioning that the transitions of rate $\lambda_2$ in states $q_l=2$ and $q_l=3$ are omitted as these transitions will not induce a transition between the discrete states nor any reset on the age process and are therefore irrelevant}. Moreover, since user $1$ has currently no available packets, we set the age of its packet $x_1$ to $0$ after the transition since $x_1$ is already $0$. This done only to simplify further the equations that need to be solved. The same thing happens when transition $l=5$ takes place, however there is a subtle difference: in $q_l=1$, there is already a packet for user $1$ that is being served and therefore transition $l=5$ will not set $x_1$ to $0$ but rather keeps it at its current value.
\item The transitions $l=4$ and $l=8$ corresponds to the end of the service of a packet for user $1$. The two transitions have different rates due to the fact that in state $q_l=2$, NOMA is applied between user $1$ and user $2$. In both cases, the age process of user $1$ at the monitor will experience a drop to $x_1$ while the age of the packet in the system of user $1$, $x_1$, is set to $0$ as the system is now empty.
\item The transitions $l=6$ and $l=10$ happen when a packet for user $2$ finishes being served. Similarly, the two transitions have different rates due to the fact that in state $q_l=2$, NOMA is applied between the two users. Both transitions will not have an effect on the age process of user $1$ at the monitor. As it has been previously done, the subtle difference is that the transition $l=10$ sets the age of the packet for user $1$, $x_1$, to $0$ since user $1$ has currently no packets available.
\end{enumerate}
\begin{figure}[!ht]
\centering
\includegraphics[width=.60\linewidth]{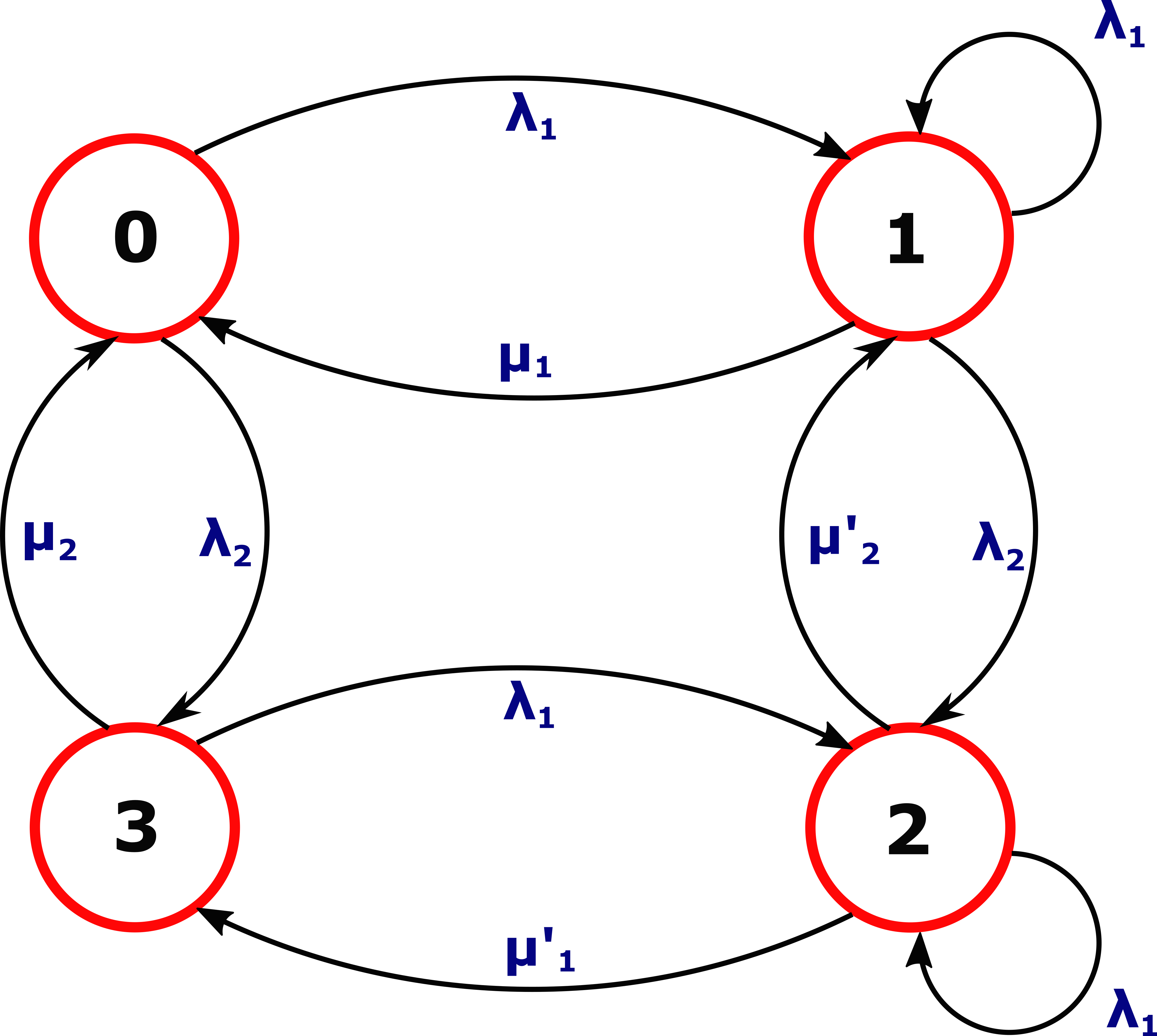}
\setlength{\belowcaptionskip}{-10pt}
\caption{Illustration of the NOMA stochastic hybrid systems Markov chain}
\label{effect}
\end{figure}

%

Next, we investigate the differential equations dictating the evolution of the age process in each discrete state. Based on the description of the system, we assert that the age at the monitor $x_0$ is always increasing at a unit rate. On the other hand, the age process of the packet of user $1$ increases at a unit rate when there is one in its system. More specifically: 
\begin{equation*}
\boldsymbol{b}_q=[1\:\:0] \quad q\in \{0,3\}
\end{equation*}
\begin{equation}
\boldsymbol{b}_q=[1\:\:1] \quad q\in \{1,2\}
\label{bstuff1}
\end{equation}
Next, to apply Theorem \ref{theomhem}, we have to find the stationary distribution of the Markov chain $q(t)$, denoted by $\overline{\boldsymbol{\pi}}$. To do so, it is sufficient to formulate the global balance equations at states $\{0,2,3\}$ and take into account that $\sum_{q\in\mathbb{Q}}\overline{\pi}_q=1$. The exact expression of the stationary distribution can be therefore easily formulated by invoking any standard method of solving the system of linear equations in (\ref{eqlin}). Finally, by applying Theorem \ref{theomhem}, and by taking into account the transitions of Table \ref{trans} and the differential equations vector of (\ref{bstuff1}), we can show that $v_{01}=v_{31}=0$. This is due to the fact that in these states, there is not a packet in the system of user $1$. Moreover, it can be shown that the vector $\boldsymbol{\overline{v}}'$ verify the linear system reported in (\ref{eqlin2}). The exact expression of the average age of user $1$ can be therefore concluded easily by invoking any standard method of solving the system of linear equations in (\ref{eqlin2}).
\end{IEEEproof}
\subsection{Conventional OMA}
In this section, we forgo the ability of employing the NOMA scheme. More specifically, as there are two users to serve, once a packet arrives to either of the users, this user is immediately served. Once the user has been served, the other user's packets starts being served if there are any. Our goal is to calculate the total average age of the network in this scenario. Similarly as the previous section and for the same reasons, we examine the network solely from the perspective of user $1$. For this purpose, we provide the following theorem.
\begin{theorem}
In the aforementioned OMA settings, the average age of user $1$ is equal to $\overline{\Delta}_1=\sum\limits_{q\in\mathbb{Q}}\overline{v}_{q0}=\overline{v}'_1+\overline{v}'_2+\overline{v}'_4+\overline{v}'_5+\overline{v}'_7$ 
with $\boldsymbol{\overline{v}}'=[\overline{v}_{00},\overline{v}_{10},\overline{v}_{11},\overline{v}_{20},\overline{v}_{30},\overline{v}_{31},\overline{v}_{40},\overline{v}_{41}]^{T}$ being the solution to:
\begin{equation}
\boldsymbol{A}_2\boldsymbol{\overline{v}'}=\boldsymbol{c}_2
\label{eqlin6}
\end{equation}
where:
\begin{equation}
\boldsymbol{A}_2=
\begin{psmallmatrix}
    \lambda_1+\lambda_2 & 0 &-\mu_1 &-\mu_2 & 0 & 0  & 0 & 0  \\
    -\lambda_1 & \lambda_2+\mu_1 &0 &0 & -\mu_2 & 0  & 0 & 0  \\
     0 & 0 &\lambda_1+\lambda_2+\mu_1 &0 & 0 & -\mu_2  & 0 & 0  \\
   -\lambda_2 & 0 &0 &\lambda_1+\mu_2 & 0 & 0  & 0 & -\mu_1  \\
      0 & 0 &0 &-\lambda_1 & \mu_2 & 0  & 0 & 0  \\
     0 & 0 &0 &0 & 0 & \lambda_1+\mu_2  & 0 & 0  \\
     0 & -\lambda_2 &0 &0 & 0 & 0  & \mu_1 & 0  \\
     0 & 0 &-\lambda_2 &0 & 0 & 0  & 0 & \lambda_1+\mu_1  \\
\end{psmallmatrix}
\end{equation}
\begin{equation}
\boldsymbol{c}_2=\begin{psmallmatrix}
    \overline{\pi}_0  \\
    \overline{\pi}_1  \\
\overline{\pi}_1  \\
\overline{\pi}_2 \\
\overline{\pi}_3   \\
\overline{\pi}_3 \\
\overline{\pi}_4   \\
\overline{\pi}_4 
\end{psmallmatrix}
\end{equation}
and $\overline{\boldsymbol{\pi}}$ is the stationary distribution of the Markov chain $q(t)$ that can be found by solving:
\begin{equation}
\boldsymbol{A}_1\boldsymbol{\overline{\pi}}=\boldsymbol{c}_1
\label{eqlin4}
\end{equation}
where:
\begin{equation}
\boldsymbol{A}_1=
\begin{psmallmatrix}
    \lambda_1+\lambda_2 & -\mu_1 & -\mu_2 & 0 & 0\\
    -\lambda_1 & \mu_1+\lambda_2 & 0 & -\mu_2 & 0\\
    -\lambda_2 & 0 & \lambda_1+\mu_2 & 0 & -\mu_1\\
    0 & 0 & -\lambda_1 & \mu_2 & 0\\
    1 & 1 & 1 & 1 & 1
\end{psmallmatrix}
\end{equation}
\begin{equation}
\boldsymbol{c}_1=\begin{psmallmatrix}
    0  \\
    0  \\
    0  \\
0  \\
1  \\
\end{psmallmatrix}
\end{equation}
\label{averageageoma}
\end{theorem}
\begin{IEEEproof} We first start by defining the discrete states set $\mathbb{Q}=\{0,1,2,3,4\}$ where:
\begin{enumerate}
\item $q(t)=0$: no packets are available for either of the two users and therefore the server is idle at time $t$
\item $q(t)=1$: user $1$ is being served while user $2$ has no packet at time $t$
\item $q(t)=2$: user $2$ is being served while user $1$ has no packet at time $t$
\item $q(t)=3$: user $2$ is being served while user $1$ is waiting to be served at time $t$
\item $q(t)=4$: user $1$ is being served while user $2$ is waiting to be served at time $t$
\end{enumerate}
The continuous-time state process is defined as $\boldsymbol{x}(t)=[x_{0}(t),x_{1}(t)]$ where $x_{0}(t)$ is the age of user $1$ at the monitor at time $t$ and $x_{1}(t)$ is the age of the packet in the system of user $1$ at time $t$. We present a summary of the transitions between the discrete states and the reset maps they induce on the age process $\boldsymbol{x}(t)$ in the following table:
\begin{center}
\begin{tabular}{cccccc}
$l$ & $q_l\rightarrow q'_{l}$ & $\lambda^{(l)}$ & $\boldsymbol{xA_l}$ & $\boldsymbol{A_l}$  &  $\boldsymbol{v}_{q_l}\boldsymbol{A_l}$ \\
\hline
 $1$ & $0\rightarrow1$  & $\lambda_1$ & $[x_0,0]$ & $\begin{psmallmatrix}
    1 & 0  \\
    0& 0
\end{psmallmatrix}$  &  $[v_{00},0]$\\
 $2$ & $0\rightarrow2$  & $\lambda_2$ & $[x_0,0]$ & $\begin{psmallmatrix}
    1 & 0  \\
    0& 0
\end{psmallmatrix}$  &  $[v_{00},0]$\\
 $3$ & $1\rightarrow0$  & $\mu_1$ & $[x_1,0]$ & $\begin{psmallmatrix}
    1 & 0  \\
    0& 0
\end{psmallmatrix}$  &  $[v_{11},0]$\\
 $4$ & $1\rightarrow1$  & $\lambda_1$ & $[x_0,0]$ & $\begin{psmallmatrix}
    1 & 0  \\
    0& 0
\end{psmallmatrix}$  &  $[v_{10},0]$\\
 $5$ & $1\rightarrow2$  & $\lambda_2$ & $[x_0,x_1]$ & $\begin{psmallmatrix}
    1 & 0  \\
    0& 1
\end{psmallmatrix}$  &  $[v_{10},v_{11}]$\\
 $6$ & $2\rightarrow0$  & $\mu_2$ & $[x_0,x_1]$ & $\begin{psmallmatrix}
    1 & 0  \\
    0& 1
\end{psmallmatrix}$  &  $[v_{20},v_{21}]$\\
$7$ & $2\rightarrow3$  & $\lambda_1$ & $[x_0,0]$ & $\begin{psmallmatrix}
    1 & 0  \\
    0& 0
\end{psmallmatrix}$  &  $[v_{20},0]$\\
 $8$ & $3\rightarrow3$  & $\lambda_1$ & $[x_0,0]$ & $\begin{psmallmatrix}
    1 & 0  \\
    0& 0
\end{psmallmatrix}$  &  $[v_{30},0]$\\
 $9$ & $3\rightarrow1$  & $\mu_2$ & $[x_0,x_1]$ & $\begin{psmallmatrix}
    1 & 0  \\
    0& 1
\end{psmallmatrix}$  &  $[v_{30},v_{31}]$\\
 $10$ & $4\rightarrow4$  & $\lambda_1$ & $[x_0,0]$ & $\begin{psmallmatrix}
    1 & 0  \\
    0& 0
\end{psmallmatrix}$  &  $[v_{40},0]$\\
$11$ & $4\rightarrow2$  & $\mu_1$ & $[x_1,0]$ & $\begin{psmallmatrix}
    0 & 0  \\
    1& 0
\end{psmallmatrix}$  &  $[v_{41},0]$\\
\end{tabular}
 \captionof{table}{Stochastic Hybrid System Description}
 \label{trans2}
\end{center}
In the following, we elaborate on the transitions reported in Table \ref{trans}:
\begin{enumerate}
\item The set of transitions of rate $\lambda^{(l)}=\lambda_1$ corresponds to the case where a new packet arrives to user $1$. This arrival can occur when user $1$ has no available packets ($l=1$ and $l=7$). It can also replace a previous packet in user $1$'s system ($l=8$) or preempts a packet that is already in service ($l=4$ and $l=10$ ). In all cases, this transition will not influence the age process $x_0$ at the monitor but will reset the age of the packet of user $1$ to $0$.
\item The set of transitions of rate $\lambda^{(l)}=\lambda_2$ corresponds to the case where a new packet arrives to user $2$. These transitions will have no effect on the age of user $1$ at the monitor. Moreover, in the states where there is not a packet in user $1$'s system, these transitions are chosen to induce a reset $x_1=0$.
\item The set of transitions of rate $\lambda^{(l)}=\mu_1$ corresponds to the case where a packet of user $1$ finished being served. In this case, the age process of user $1$ at the monitor will experience a drop to $x_1$ while the age of the packet $x_1$ is set to $0$ as the system is now empty.
\item The set of transitions of rate $\lambda^{(l)}=\mu_2$ corresponds to the case where a packet of user $2$ finishes being served. These transitions will have no effect on the age of user $1$ at the monitor. Moreover, in the states where there is not a packet in user $1$'s system, these transitions are chosen to induce a reset $x_1=0$.
%
\end{enumerate}
\begin{figure}[!ht]
\centering
\includegraphics[width=.85\linewidth]{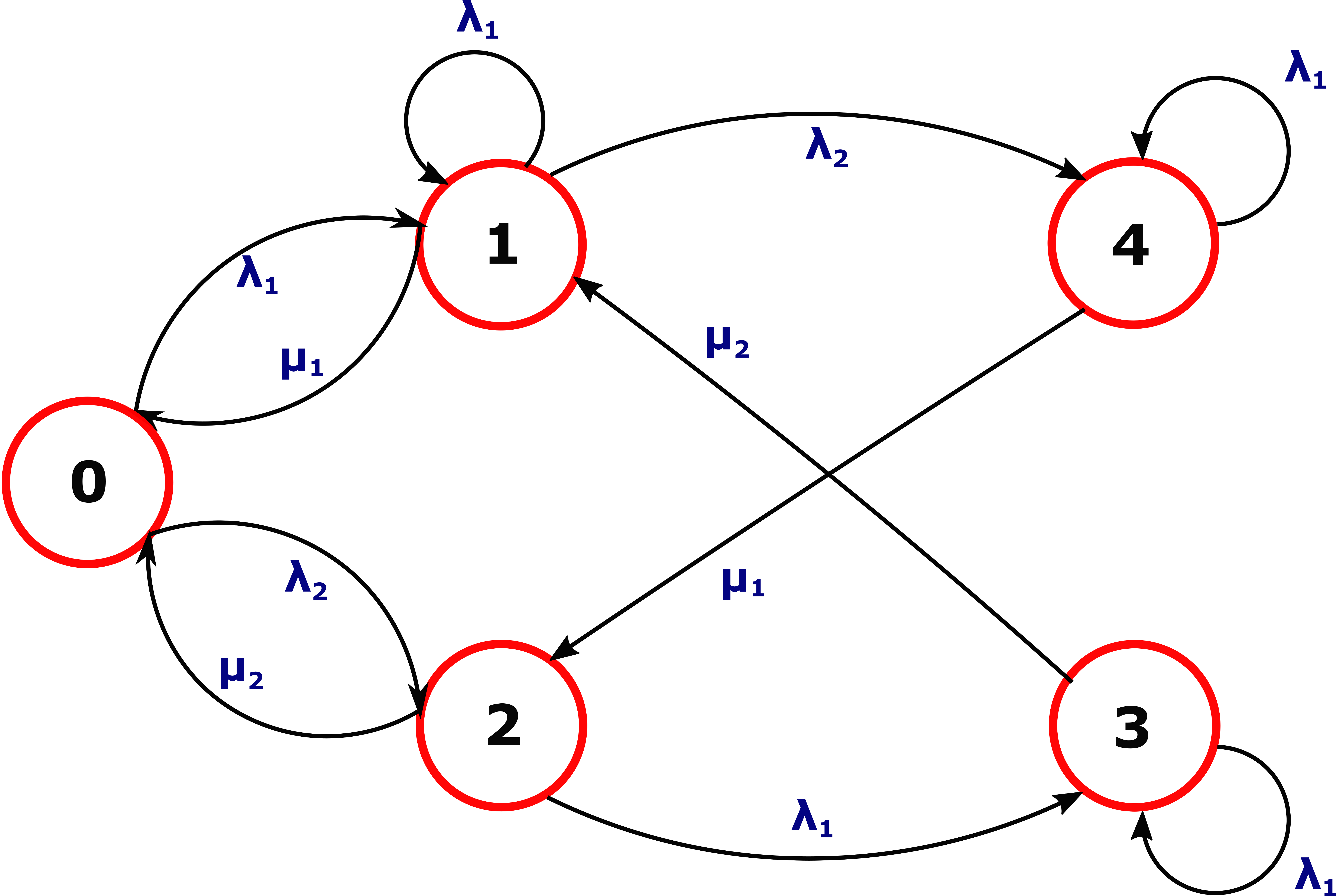}
\setlength{\belowcaptionskip}{-10pt}
\caption{Illustration of the OMA stochastic hybrid systems Markov chain}
\label{secondchain}
\end{figure}
We now tackle the differential equations governing the age process in each state. Similarly as the previous section, the age at the monitor $x_0$ is always increasing at a unit rate. When the system of user $1$ is not empty, the age process $x_1$ increases at a unit rate. Consequently, we have:
\begin{equation*}
\boldsymbol{b}_q=[1\:\:0] \quad q\in \{0,2\}
\end{equation*}
\begin{equation}
\boldsymbol{b}_q=[1\:\:1] \quad q\in \{1,3,4\}
\label{bstuff2}
\end{equation}
As previously done in Theorem \ref{averageagenoma}, we then seek the stationary distribution of the Markov chain reported in Fig. \ref{secondchain} by formulating the global balance equations. Then, we can apply Theorem \ref{theomhem} to conclude the proof.
\end{IEEEproof}
\section{Numerical Results}
This section provides numerical results in the aim of giving insights based on the theoretical analysis provided in the previous section. In the first scenario, we suppose that $\lambda_1,\lambda_2\rightarrow+\infty$ which corresponds to the case where users always have fresh data to send and packets arrive fresh to the receiver. The motivation for this scenario is to eliminate the effect of the arrival process on the age process in order to fully highlight the difference between NOMA and OMA. We consider that the average service rate of each user is $\mu_1=1$ and $\mu_2=2$. When OMA is employed and with $\lambda_1,\lambda_2\rightarrow+\infty$, one can easily see that the evolution of the system becomes a Round Robin one where users take turn in their transmission (sharing factor $\delta=0.5$). In this case, the average number of packets sent per unit of time is $\mu_t=\frac{\mu_1+\mu_2}{2}=1.5$. By applying Theorem \ref{averageageoma}, we end up with:
\begin{equation}
\overline{\Delta}_{OMA}=\frac{1}{\mu_1}+\frac{1}{\mu_2}+\frac{\frac{1}{\mu_1}}{1+\frac{\mu_1}{\mu_2}}+\frac{\frac{1}{\mu_2}}{1+\frac{\mu_2}{\mu_1}}
\end{equation}
On the other hand, when NOMA is employed and with $\lambda_1,\lambda_2\rightarrow+\infty$, both users will be sending packets simultaneously at all time. Due to the spectral efficiency improvement of NOMA, we assume that the average number of packets sent per unit of time is $\mu'_t=\alpha\mu_t$ where $\alpha \:\in [1,2]$ can be seen as the spectral efficiency multiplication factor. More specifically, $\alpha=1$ corresponds to the case where NOMA gives no spectral efficiency improvement in comparison to OMA while $\alpha=2$ corresponds to the extreme case where the interference between the users can be completely eliminated. Consequently, we consider that $\mu'_1=\alpha\frac{\mu_1}{2}$ and $\mu'_2=\alpha\frac{\mu_2}{2}$ (we recall that the sharing factor $\delta$ is equal to $0.5$). The average age can easily be verified to be using Theorem \ref{averageagenoma}: 
\begin{equation}
\overline{\Delta}_{NOMA}=\frac{1}{\mu'_1}+\frac{1}{\mu'_2}
\end{equation}
\begin{figure}[!ht]
\centering
\includegraphics[width=.75\linewidth]{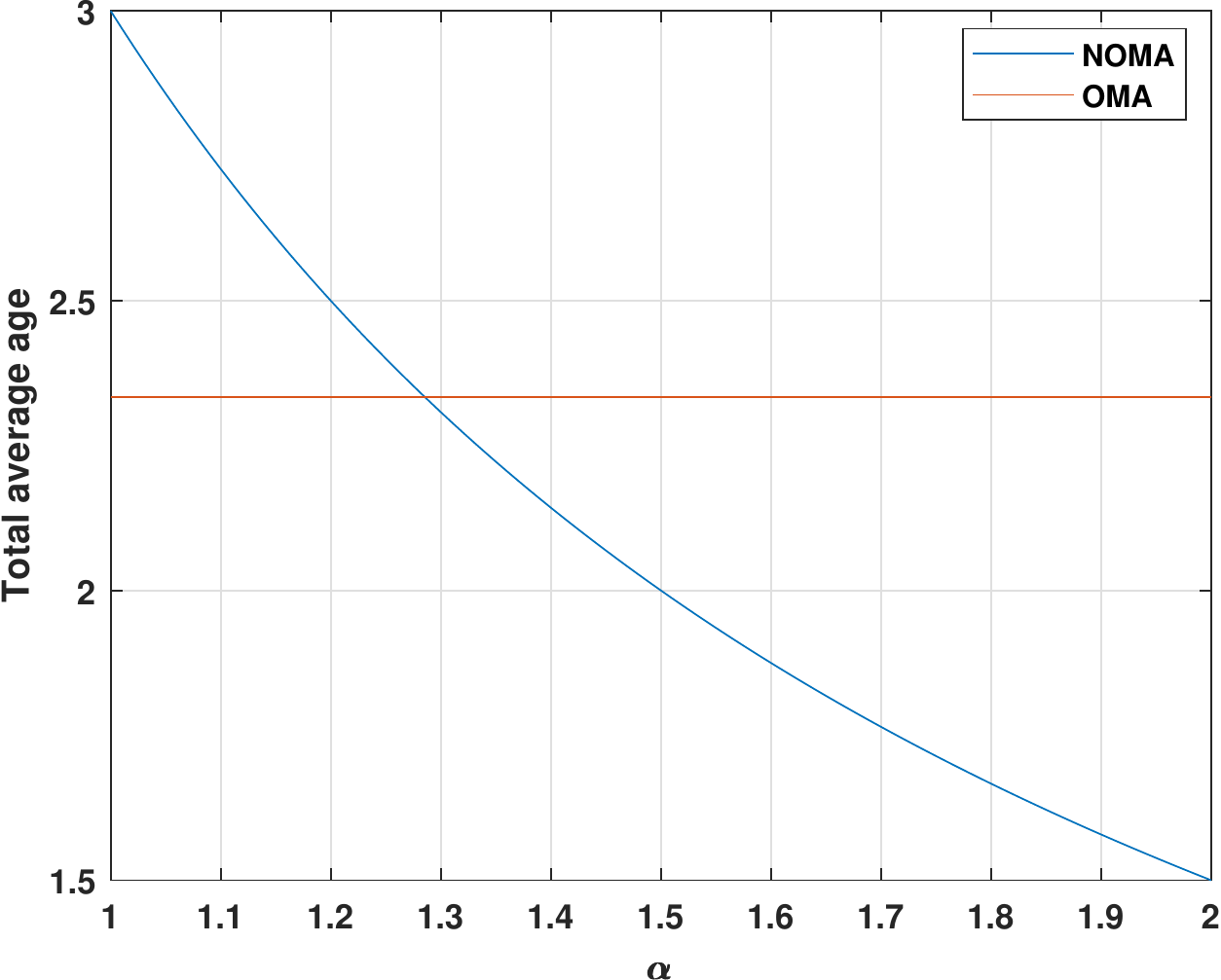}
\setlength{\belowcaptionskip}{-10pt}
\caption{Illustration of the average age in OMA and NOMA environments}
\label{nomaomaresult}
\end{figure}\\
One can see in Fig. \ref{nomaomaresult} that a spectral multiplication factor $\alpha>1$ does not necessarily translate into a smaller average age. In fact, for $\alpha=1.2$ (i.e. $20\%$ more packets are being sent by the users), OMA still achieves better age performance than its NOMA counterpart. This is a consequence of the key difference between data communication systems and status update systems: when a packet of user $1$ is successfully transmitted, the age of this user drops. An update from user $1$ right after does not carry as much fresh information since the monitor has just received an update on the process that is being monitored. Therefore, it makes sense to let user $2$ takes the channel solely for itself at a faster rate to send its much needed information to the monitor. On the NOMA counterpart, both users keep transmitting at the same time, with a reduced individual rates ($\mu'_1<\mu_1$ and $\mu'_2<\mu_2$) but with a higher sum rate ($\mu'_t=\alpha\mu_t$). In other words, even when user $1$ has just sent an update, this user is still granted the channel in attempt to increase the spectral efficiency of the network. Another thing we notice is that as the multiplication factor grows higher, this gap reduces until NOMA outperforms OMA. These results suggest that when the increase in spectral efficiency is really high in comparison to OMA, NOMA has the potential of outperforming OMA.

In the second scenario, we consider that the service rate is $(\mu_1=0.1$, $\mu_2=0.2)$ and we set $\alpha=1.2$. We examine the gap between NOMA and OMA in function of the arrival rate ($\lambda_1=\lambda_2=\lambda$). We can see that for really small arrival rate, the two schemes virtually coincide in performance. This is due to the fact that when the arrival rate is really small, with a high probability, each user will be transmitting alone on the orthogonal block. However, as the arrival rate grows higher, a gap to OMA starts showing. This is due to the fact that the orthogonal block starts being shared more often between the two users. All these results suggest that NOMA should be carefully investigated in the case of minimizing the AoI of the network and should not be taken for granted as a definite performance improvement due to its promised increase of spectral efficiency.
\begin{figure}[!ht]
\centering
\includegraphics[width=.75\linewidth]{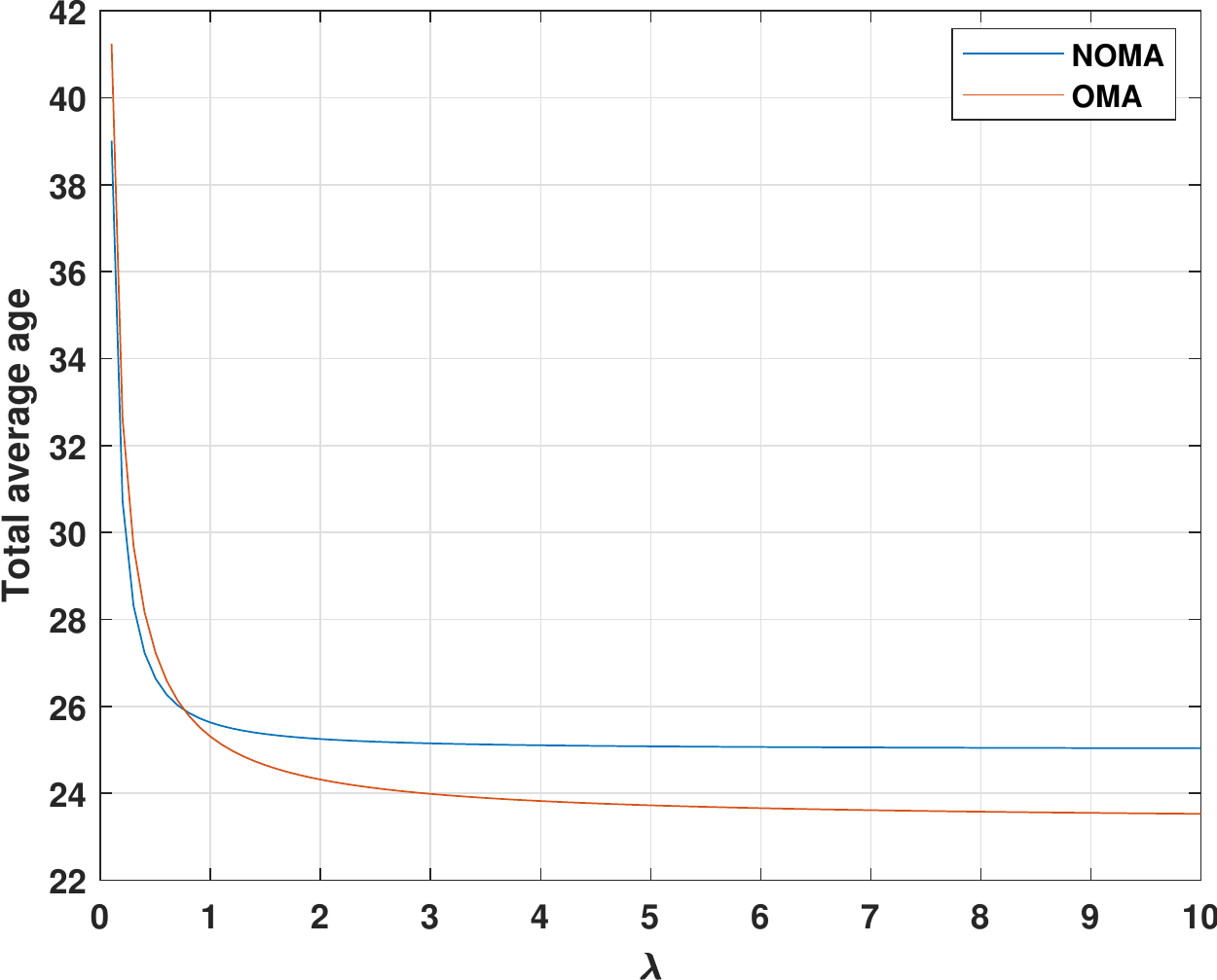}
\setlength{\belowcaptionskip}{-10pt}
\caption{Illustration of the average age in OMA and NOMA environments in function of the arrival rate}
\label{nomaomaresult}
\end{figure}

\section{Conclusion}
In this paper, we have investigated the age of information in the framework of NOMA, a technology rivaling OMA in 3GPP standardization for future 5G networks machine type communications. By leveraging the notion of stochastic hybrid systems, we have found a closed form of the average age of the network in both NOMA and conventional OMA environments. By studying the average age in both cases, we drew a comparison between the two schemes in terms of the total average AoI in the network. Surprisingly, it was shown that even when NOMA achieves better spectral efficiency in comparison to OMA, this does not necessarily translates into a lower average AoI in the network.
\bibliographystyle{IEEEtran}
\bibliography{trialout}

\end{document}